\documentclass[aps,prl,twocolumn,superscriptaddress,floatfix,10pt]{revtex4-1}
\usepackage{graphicx}
\usepackage{dcolumn}
\usepackage{bm}
\usepackage{amsfonts}
\usepackage{amssymb}
\usepackage{amsmath}
\usepackage{fix-cm}
\usepackage{mathptmx} 
\usepackage[T1]{fontenc}
\usepackage[colorlinks,allcolors=blue]{hyperref}
\makeatletter
\def\NAT@def@citea{\def\@citea{\NAT@separator}}
\makeatother

\begin{document}

\title{Dependence of fusion on isospin dynamics}

\author{K. Godbey}\email{kyle.s.godbey@vanderbilt.edu}
\affiliation{Department of Physics and Astronomy, Vanderbilt University, Nashville, TN 37235, USA}

\author{A.S. Umar}\email{umar@compsci.cas.vanderbilt.edu}
\affiliation{Department of Physics and Astronomy, Vanderbilt University, Nashville, TN 37235, USA}

\author{C. Simenel}\email{cedric.simenel@anu.edu.au}
\affiliation{Department of Nuclear Physics, Research School of Physics and Engineering, The Australian National University, Canberra ACT  2601, Australia}

\date{\today}

\begin{abstract}
We introduce a new microscopic approach to calculate the dependence of fusion barriers and cross-sections on
isospin dynamics. The method is based on the time-dependent Hartree-Fock theory and
the isoscalar and isovector properties of the energy density functional (EDF). The contribution to
the fusion barriers originating from the isoscalar and isovector parts of the EDF is calculated.
It is shown that for non-symmetric systems the isovector dynamics influence the sub-barrier fusion
cross-sections. For most systems this results in an enhancement of the sub-barrier cross-sections,
while for others we observe differing degrees of hindrance.
We use this approach to provide an explanation of recently measured fusion cross sections which show a surprising enhancement at low $E_\mathrm{c.m.}$ energies
for the system $^{40}$Ca+$^{132}$Sn as compared to the more neutron-rich system
$^{48}$Ca+$^{132}$Sn, and discuss the dependence of sub-barrier fusion cross-sections on transfer.

\end{abstract}
\maketitle

One of the major open questions in fusion reactions of exotic neutron-rich nuclei
is the dependence of the fusion cross section on the neutron excess,
or equivalently on the total isospin quantum number $T_z = (Z-N)/2$.
This is a timely subject given the expected
availability of increasingly exotic beams at rare isotope facilities\,\cite{balantekin2014}.
The influence of isospin dynamics on fusion is also one of the key questions pertaining to the
production of superheavy elements using neutron rich nuclei\,\cite{loveland2007}.
Besides being a fundamental nuclear structure and reaction question, the answer to this inquiry
is also vital to our understanding of the nuclear equation of state (EOS) and symmetry energy\,\cite{li2014}. 
The EOS plays a key role in elucidating the structure of exotic nuclei\,\cite{chen2014},
the dynamics of heavy ion collisions\,\cite{danielewicz2002,tsang2009},
the composition of neutron stars\,\cite{haensel1990,chamel2008,horowitz2004,utama2016}, and the mechanism of core-collapse supernovae\,\cite{bonche1981,watanabe2009,shen2011}.
The influence of isospin flow during heavy-ion reaction is usually discussed in term of the
$(N/Z)$ asymmetry of the target and projectile or the $Q$-values for neutron transfer.
Dynamically, $(N/Z)$ asymmetry leads to the so-called pre-equilibrium giant-dipole resonance 
(GDR), which has been studied using various dynamical 
approaches\,\cite{dasso1985,chomaz1993,baran1996,baran2009,baran2005,baran2001,simenel2001,simenel2007,sekizawa2013}.
The main issue with these microscopic calculations is that it is often difficult, if not impossible, to quantitatively assess the
influence of the GDR on fusion cross-sections.
Other approaches have also been considered to study the impact of neutron transfer on fusion barriers and fusion cross-sections.
These include the coupled-channels (CC) approach\,\cite{rowley1992,hagino2012,bourgin2016}
and models incorporating intermediate neutron 
rearrangements\,\cite{zagrebaev2003,zagrebaev2007c,karpov2015}.

The predictions of fusion enhancement based on $Q$-value arguments have been recently challenged
by a series of experiments carried out
with radioactive $^{132}$Sn beams and with stable $^{124}$Sn beams on
$^{40,48}$Ca targets~\cite{kolata2012}. It turns out that the $^{40}$Ca+Sn systems
have many positive $Q$-values for neutron-pickup while all the $Q$-values for
$^{48}$Ca+Sn are negative. However, the data analysis reveals that the fusion
enhancement is not proportional to the magnitudes of those $Q$-values.
Particularly puzzling is the experimental observation of a sub-barrier fusion
enhancement in the system $^{40}$Ca+$^{132}$Sn
as compared to the more neutron-rich  $^{48}$Ca+$^{132}$Sn system.
Part of the puzzle comes from the fact that the $8$ additional neutrons in
$^{48}$Ca should increase the attractive strong nuclear interaction
and thus lower the fusion barrier, resulting in an enhanced sub-barrier
fusion cross section.
One explanation is that the dynamics will ``wash out'' the static structure effects such as the
Coulomb barrier lowering due to neutron skin\,\cite{vophuoc2016}.
This anomaly was also studied in a recent experiment\,\cite{liang2016}.

In this work we address the impact of isospin dynamics on fusion barriers and cross-sections using
the microscopic
time-dependent Hartree-Fock (TDHF) theory\,\cite{negele1982,simenel2012} 
together with the density-constrained TDHF (DC-TDHF) method for calculating fusion barriers\,\cite{umar2006b}.
In the TDHF approximation the many-body wavefunction is replaced by a single
Slater determinant and this form is preserved at all times, implying that two-body correlations
are neglected.
In this limit, the
variation of the time-dependent action with respect to the single-particle states, $\phi^{*}_{\lambda}$, yields the most probable time-dependent path
in the multi-dimensional space-time phase space
represented as a
set of coupled, nonlinear, self-consistent initial value equations
for the single-particle states
\begin{equation}
h(\{\phi_{\mu}\}) \ \phi_{\lambda} (r,t) = i \hbar \frac{\partial}{\partial t} \phi_{\lambda} (r,t)
\ \ \ \ (\lambda = 1,...,A)\,,
\label{eq:TDHF}
\end{equation}
where $h$ is the HF single-particle Hamiltonian.
These are the fully microscopic time-dependent Hartree-Fock equations.

Almost all TDHF calculations employ the Skyrme EDF, which allows the total energy of the system to be represented
as an integral of the energy density ${\cal H}(\mathbf{r})$\,\cite{engel1975}
\begin{equation}
\label{eq:energy}
E = \int d^3\mathbf{r} {\cal H}(\mathbf{r})\,,
\end{equation}
which includes the kinetic,
isoscalar, isovector, and Coulomb terms \,\cite{dobaczewski1995}:
\begin{equation}
\label{eq:edensity}
{\cal H}(\mathbf{r}) = \frac{\hbar^2}{2m}\tau_0
+ {\cal H}_0(\mathbf{r})
+ {\cal H}_1(\mathbf{r})
+ {\cal H}_C(\mathbf{r})\,.
\end{equation}
In particular, 
\begin{widetext}
\begin{equation}
\label{eq:efunctional}
{\cal H}_\mathrm{I}(\mathbf{r})
= C_\mathrm{I}^{\rho}            \rho_\mathrm{I}^2
+  C_\mathrm{I}^{   s}            \mathbf{s}_\mathrm{I}^2
+  C_\mathrm{I}^{\Delta\rho}      \rho_\mathrm{I}\Delta\rho_\mathrm{I}
+  C_\mathrm{I}^{\Delta s}        \mathbf{s}_\mathrm{I}\cdot\Delta\mathbf{s}_\mathrm{I}
+  C_\mathrm{I}^{\tau}      \left(\rho_\mathrm{I}\tau_\mathrm{I}-\mathbf{j}_\mathrm{I}^2  \right)
+  C_\mathrm{I}^{   T}      \Big(\mathbf{s}_\mathrm{I}\cdot
\mathbf{T}_\mathrm{I} - \tensor{J}_\mathrm{I}^2\Big)
+ C_\mathrm{I}^{\nabla J}  \Big(\rho_\mathrm{I}\mathbf{\nabla}\cdot\mathbf{J}_\mathrm{I}
+ \mathbf{s}_\mathrm{I}\cdot
(\mathbf{\nabla}\times\mathbf{j}_\mathrm{I})\Big)\,,
\end{equation}
\end{widetext}
where we have used the gauge invariant form suitable for time-dependent calculations.
The isospin index $\mathrm{I}=0,1$ for isoscalar and isovector energy densities, respectively. 
The most common choice of Skyrme EDF restricts the density dependence of the
coupling constants to the $C_\mathrm{I}^{\rho}$ and $C_\mathrm{I}^s$ terms only.
These density dependent coefficients induce the coupling of isoscalar and isovector fields
in the Hartree-Fock Hamiltonian.
The isoscalar (isovector) energy density, ${\cal H}_0(\mathbf{r})$ (${\cal H}_1(\mathbf{r})$), depends on the isoscalar (isovector) particle
density, $\rho_0 = \rho_n + \rho_p$ ($\rho_1 = \rho_n - \rho_p$), with analogous expressions for other densities and
currents. 
Values of the coupling
coefficients as well as their relation to the alternative parametrizations of the Skyrme
EDF can be found in\,\cite{dobaczewski1995}.
\begin{figure}[!htb]
	\includegraphics*[width=8.0cm]{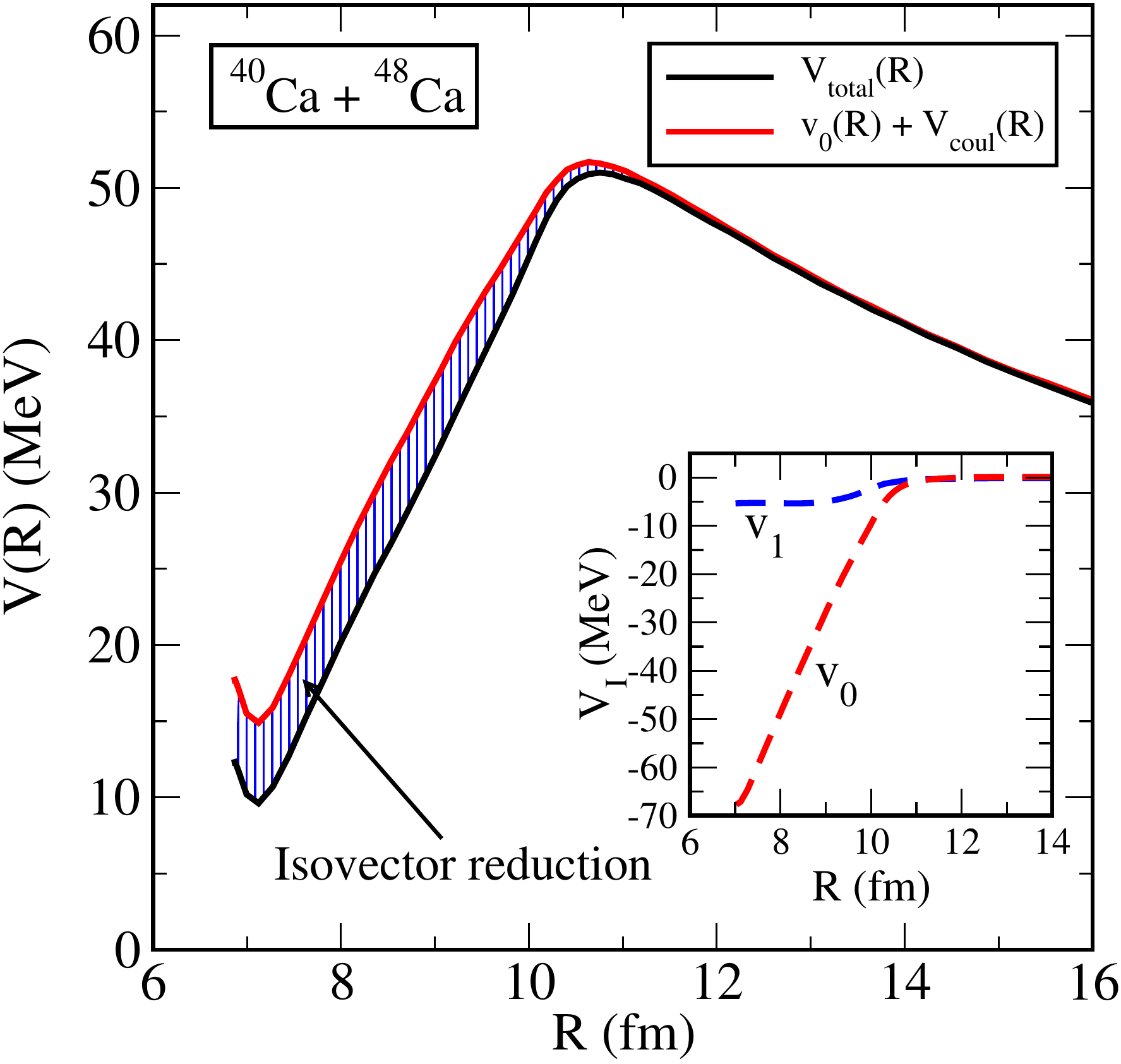}
	\caption{(Color online) For the $^{40}$Ca+$^{48}$Ca system;
		Total and isoscalar DC-TDHF potentials. The shaded region
		corresponds to the reduction originating from the isovector contribution to the energy
		density. The insert shows the isoscalar and isovector contributions to the interaction barrier
		without the Coulomb potential.
	    The TDHF collision energy was $E_\mathrm{c.m.}=55$\,MeV.}
	\label{fig:CaCa}
\end{figure}

The above form of the EDF is more suitable for studying the isospin dependence
of nuclear properties and have been employed in nuclear structure studies\,\cite{dobaczewski1995}.
In the same spirit we can utilize this approach to study isospin dependent effects in nuclear
reactions microscopically.
In particular, the density-constrained time-dependent Hartree-Fock (DC-TDHF) method\,\cite{umar2006b} can be
employed to study isospin effects on fusion barriers and fusion cross-sections.
The DC-TDHF approach calculates the nucleus-nucleus potentials $V(R)$ directly 
from  TDHF dynamics and has been used to calculate fusion cross-sections for a wide range of
reactions\,\cite{umar2014a,simenel2013a,umar2012a,umar2006a,oberacker2010,umar2009a,jiang2015a}.
The basic idea of this approach is the following:
At certain times $t$ or, equivalently, at certain internuclear distances
$R(t)$, a static energy minimization
is performed while constraining the proton and neutron densities to be equal to the instantaneous
TDHF densities.
We refer to the minimized energy as the ``density constrained energy''
$E_{\mathrm{DC}}(R)$.
The ion-ion interaction potential $V(R)$ is obtained by
subtracting the constant binding energies
$E_{\mathrm{A_{1}}}$ and $E_{\mathrm{A_{2}}}$ of the two individual nuclei
\begin{equation}
V(R)=E_{\mathrm{DC}}(R)-E_{\mathrm{A_{1}}}-E_{\mathrm{A_{2}}}\ .
\label{eq:vr}
\end{equation}
The calculated ion-ion interaction barriers contain all of the dynamical changes in the nuclear
density during the TDHF time-evolution in a self-consistent manner.
As a consequence of the dynamics the DC-TDHF potential is energy dependent\,\cite{umar2014a}.
Using the decomposition of the Skyrme EDF into isoscalar and isovector parts [Eq.\,(\ref{eq:efunctional})],
we can re-write this potential as
\begin{equation}
V(R) = \sum_{\mathrm{I}=0,1} v_\mathrm{I}(R) + V_C(R)\,,
\end{equation}
where $v_\mathrm{I}(R)$ denotes the potential computed by using the isoscalar and isovector parts of
the Skyrme EDF given in Eq.\,(\ref{eq:edensity}) in Eq.\,(\ref{eq:vr}).
The Coulomb potential is also calculated via Eq.\,(\ref{eq:vr}) using the Coulomb energy
density.
\begin{figure}[!htb]
	\includegraphics*[width=8.0cm]{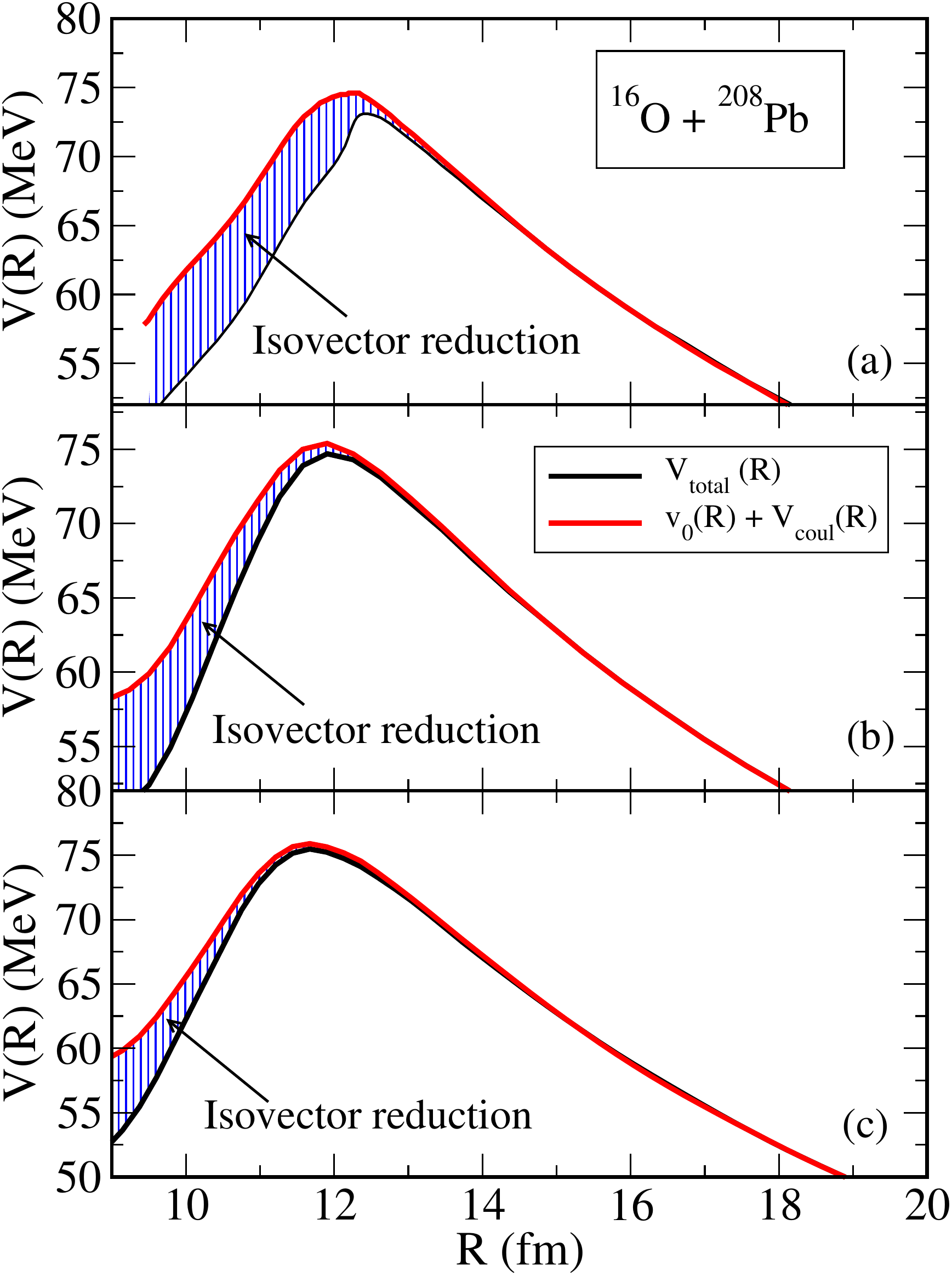}
	\caption{(Color online) For the $^{16}$O+$^{208}$Pb system;
		(a) Total and isoscalar DC-TDHF potentials at $E_\mathrm{c.m.}=75$\,MeV. The shaded
	    region corresponds to the reduction originating from the isovector contribution to
	    the energy density. 
	    (b) Same as in (a) except for $E_\mathrm{c.m.}=90$\,MeV.
        (c) Same as in (a) except for $E_\mathrm{c.m.}=120$\,MeV.}
	\label{fig:OPb}
\end{figure}

We have used the DC-TDHF approach to study fusion barriers for a number of systems.
Calculations were done in a three-dimensional
Cartesian geometry with no symmetry assumptions\,\cite{umar2006c} and using the
Skyrme SLy4 EDF\,\cite{chabanat1998a}.
The three-dimensional Poisson equation for the Coulomb potential
is solved by using Fast-Fourier Transform techniques
and the Slater approximation is used for the Coulomb exchange term.
The box size used for all the calculations
was chosen to be $60\times 30\times 30$~fm$^3$, with a mesh spacing of
$1.0$~fm in all directions. These values provide very accurate
results due to the employment of sophisticated discretization
techniques\,\cite{umar1991a}.

In Fig.\,\ref{fig:CaCa} we show the total and isoscalar fusion barriers (both including the Coulomb contribution) 
for the $^{40}$Ca+$^{48}$Ca system at $E_\mathrm{c.m.}=55$~MeV. 
 For the Ca+Ca systems the energy dependence is relatively
weak\,\cite{keser2012,umar2014a,washiyama2008}.
The reduction of the isoscalar barrier is due to the isovector contribution. It is evident that
the isovector dynamics results in the narrowing of the fusion barrier, thus resulting in an enhancement of the sub-barrier
fusion cross-sections. The insert in Fig.\,\ref{fig:CaCa} shows the isovector and isoscalar components without the Coulomb contribution.
We have also calculated fusion barriers for the $^{40}$Ca+$^{40}$Ca and $^{48}$Ca+$^{48}$Ca systems, where the isovector contribution is
zero as expected from symmetry. 

As an example of a more asymmetric system we performed calculations for the $^{16}$O+$^{208}$Pb system
at $E_{\mathrm{c.m.}}=75$\,MeV. Results are shown in Fig.\,\ref{fig:OPb}(a). Here we see a 
substantial enhancement of sub-barrier fusion due to the isovector dynamics.
For this system we have performed further
calculations at c.m. energies of 90\,MeV and 120\,MeV shown in Fig.\,\ref{fig:OPb}(b-c). 
As the beam energy  increases, the
relative contribution from the isovector component to the total barrier decreases, while the overall barrier height increases with
increasing energy. At TDHF energies much higher than the barrier height the total barriers approaches the frozen density 
barrier\,\cite{washiyama2008,umar2014a}
due to the inability of the system to rearrange at that time-scale at which time the isovector contribution vanishes as well.
\begin{figure}[!htb]
	\includegraphics*[width=8.0cm]{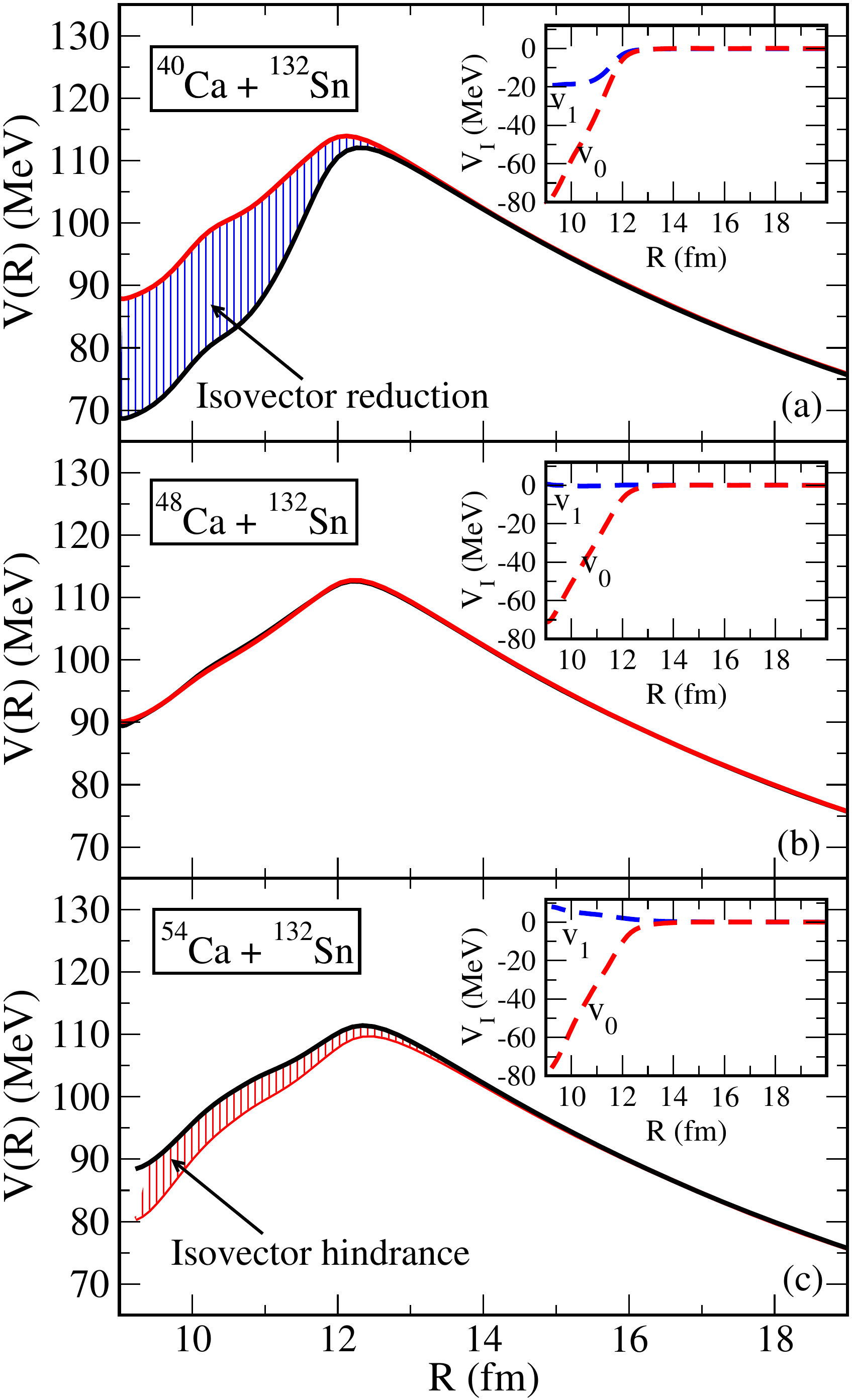}
	\caption{(Color online) For     (a) $^{40}$Ca+$^{132}$Sn,
		                            (b) $^{48}$Ca+$^{132}$Sn,
		                            (c) $^{54}$Ca+$^{132}$Sn systems;
		Total and isoscalar DC-TDHF potentials. In (a) the blue shaded region
        corresponds to the reduction originating from the isovector contribution.
        In (b) we see no isovector effect. In (c) the isovector effect is reversed
        causing hindrance as shown by the red shaded region.
		The inserts show the isoscalar and isovector contributions to the interaction barrier without the Coulomb potential.
	    The TDHF collision energy was $E_\mathrm{c.m.}=120$\,MeV.}
	\label{fig:CaSn_1}
\end{figure}
The above results demonstrate the influence of isovector dynamics on typical fusion barriers. 

We next look at Ca$+$Sn reactions. 
The experimental observation of a sub-barrier fusion enhancement in the system $^{40}$Ca+$^{132}$Sn
as compared to more neutron-rich system $^{48}$Ca+$^{132}$Sn was the subject of a previous DC-TDHF study\,\cite{oberacker2013},
where it was shown that the fusion barriers for the two systems have essentially the same height but the fusion barrier for
the $^{48}$Ca+$^{132}$Sn system was much wider than that for the $^{40}$Ca+$^{132}$Sn system.
We see in Fig.\,\ref{fig:CaSn_1}(a) a strong reduction of the isoscalar barrier due to the isovector contribution. This behavior is similar to that of the previous
two systems albeit the isovector reduction is somewhat larger as shown in the insert of
 Fig.\,\ref{fig:CaSn_1}(a).
We then performed the same calculation for the $^{48}$Ca+$^{132}$Sn system as shown in Fig.\,\ref{fig:CaSn_1}(b).
The startling result is the vanishing of the isovector contribution. With no isovector reduction the fusion barrier for this
system is much wider than that for the $^{40}$Ca+$^{132}$Sn system for which substantial reduction occurs.
The absence of the isovector component for the $^{48}$Ca+$^{132}$Sn system could be a reflection of the negative $Q-$values
for neutron pickup. This is the first direct observation of this phenomena in TDHF calculations.
This may also explain why for the $^{48}$Ca+$^{132}$Sn system simply considering the $2^+$ and
$3^-$ excitations of the target and projectile in coupled-channel calculations is able to
reproduce the sub-barrier fusion cross-sections, whereas doing the same for the $^{40}$Ca+$^{132}$Sn system grossly under-predicts the cross-sections.
In Ref.\,\cite{kolata2012}, this was attributed to the presence of significant transfer, which manifests itself in the isovector dynamics.
In Fig.\,\ref{fig:CaSn_1}(c) we have also calculated the potential barriers for the theoretical
$^{54}$Ca+$^{132}$Sn reaction. Here, we see that the influence of the isovector component
is reversed, as indicated by the shaded region. This reversal leads to the widening of
the potential barrier, further hindering sub-barrier fusion.

\begin{figure}[!hb]
	\centering
	\includegraphics[width=8.6cm]{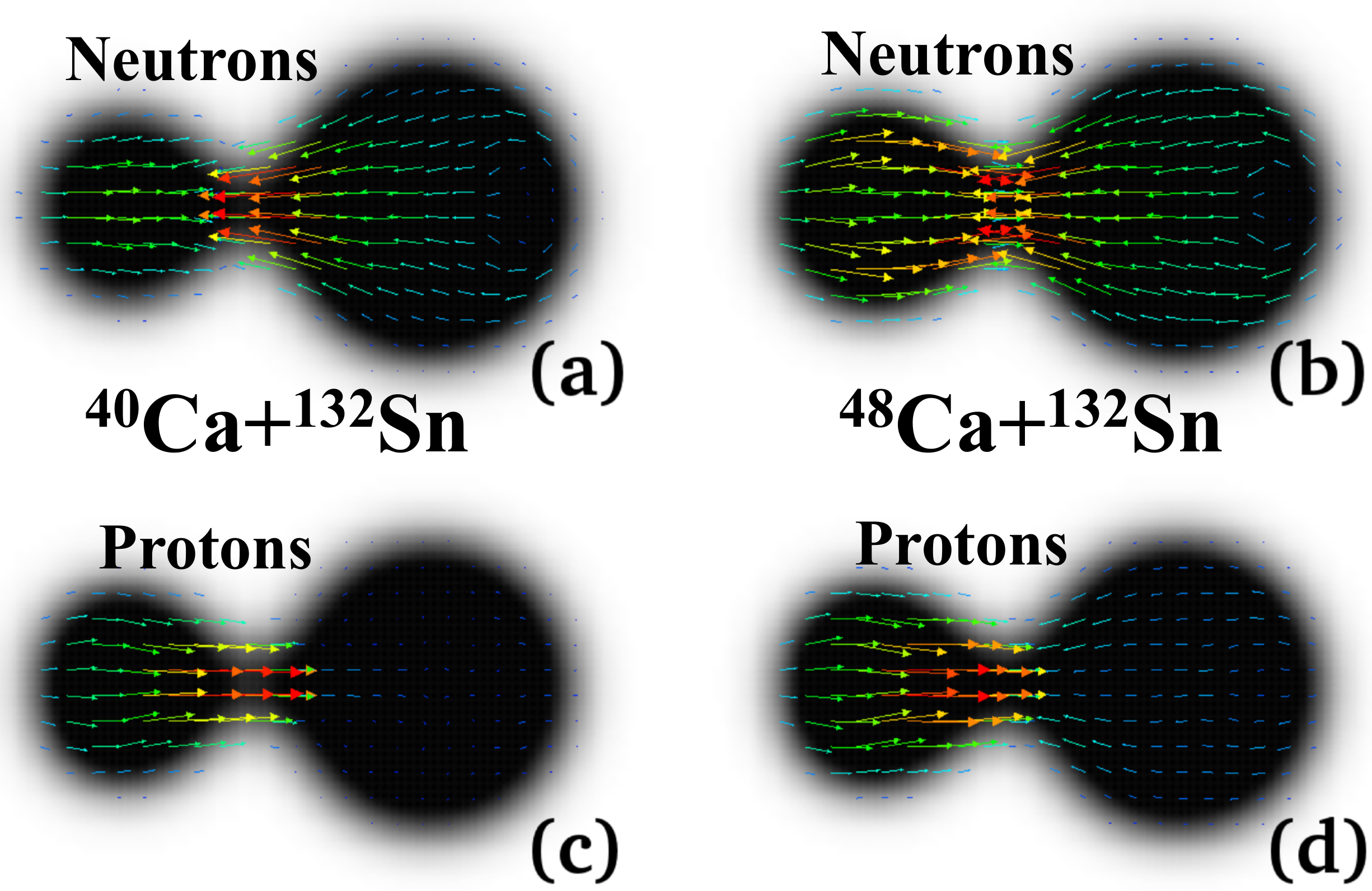}
	\caption{(Color online) Neutron and proton current vectors in $^{40,48}$Ca+$^{132}$Sn at $E_\mathrm{c.m.}=120$\,MeV and at 
		a separation  $R=11.5$\,fm between the fragments.}
	\label{fig:current}
\end{figure}

In order to investigate the role of transfer in more detail we have plotted the 
microscopic TDHF neutron and proton currents for both systems in Fig.\,\ref{fig:current},
for the same collision energy used to calculate the barriers shown in Fig.\,\ref{fig:CaSn_1} 
and at the nuclear separation $R=11.5$\,fm,
which is slightly inside the barrier but still corresponds to an early stage of the reaction. 
Examination of these currents shows that for
the $^{40}$Ca+$^{132}$Sn system neutrons flow from $^{132}$Sn to
$^{40}$Ca, whereas the proton flow is from $^{40}$Ca to $^{132}$Sn.
Proton flow for the $^{48}$Ca+$^{132}$Sn system is similar, namely from $^{48}$Ca to
$^{132}$Sn. However, we observe an interesting difference for the neutron flow in the
$^{48}$Ca+$^{132}$Sn system, namely a bi-directional flow between the two systems.
The dynamics of this bi-directional flow results in
an isovector current density in the neck region to be an order of magnitude lower
for the $^{48}$Ca+$^{132}$Sn system in comparison to the $^{40}$Ca+$^{132}$Sn.
This is the primary cause for the disappearance of the
isovector contribution to the barrier.
In the neck region, the relative magnitude of the proton current density of $^{48}$Ca+$^{132}$Sn system
is 30\%-50\% smaller than the proton current density of the $^{40}$Ca+$^{132}$Sn.
This behavior is prevalent during the entire neck dynamics.
Examination of the currents for the $^{54}$Ca+$^{132}$Sn reaction reveals that there is
essentially no through-flow of neutrons or protons during the neck formation.


In summary, we have developed a microscopic approach to study the effect of isospin dynamics on fusion barriers.
We have shown that for most systems isovector dynamics results in the thinning of the barrier thus enhancing
the sub-barrier fusion cross-sections. The isovector reduction effect vanishes for symmetric systems as well
as the $^{48}$Ca+$^{132}$Sn system for which neutron pickup $Q-$values are all negative.
These results provide a quantitative measure for the importance of transfer for
sub-barrier fusion reactions.
Furthermore, they elucidate the non-trivial dependence of sub-barrier fusion for neutron-rich
systems and illustrate the importance of dynamical microscopic models that incorporate the nuclear
structure and reactions on the same footing.
A more detailed study including cross-section ratios and other systems will be the subject of a future study.

We thank K. Vo-Phuoc for useful discussions regarding the Ca+Sn systems. 
This work has been supported by the U.S. Department of Energy under grant No.
DE-SC0013847 with Vanderbilt University and by the
Australian Research Council Grant No. FT120100760,
\bibliography{VU_bibtex_master}

\end{document}